\documentstyle[pra,aps,twocolumn,fleqn,epsfig]{revtex}

\begin{document}
\draft

\twocolumn[\hsize\textwidth\columnwidth\hsize\csname@twocolumnfalse\endcsname

\title{Entanglement in the Quantum Heisenberg $XY$ model}
\author{Xiaoguang Wang}
\address{Institute of Physics and Astronomy, Aarhus University, \\
DK-8000, Aarhus C, Denmark}
\date{\today}
\maketitle

\begin{abstract}
We study the entanglement in the quantum Heisenberg $XY$ model in which the
so-called $W$ entangled states can be generated  for 3
or 4 qubits. By the concept of concurrence, we study the entanglement
in the time evolution of the $XY$ model. We investigate the thermal entanglement 
in the two-qubit isotropic $XY$ model with a magnetic field and in the anisotropic $XY$ model, 
and find that the thermal entanglement exists for both ferromagnetic and 
antiferromagnetic cases. Some evidences of the quantum phase transition also appear in these
simple models.
\end{abstract}

\pacs{PACS numbers: 03.65.Ud, 03.67.Lx, 75.10.Jm.}
] \narrowtext

\section{Introduction}

\label{sec:intro}

Quantum entanglement has been studied intensely in recent years due to
its potential applications in quantum communication and information
processing\cite{QC} such as quantum teleportation\cite{Tele}, superdense coding\cite
{Dense}, quantum key distribution\cite{Key}, and telecoloning\cite{Clone}.
Recently D\"{u}r {\it et al.}\cite{Dur1} found that truly tripartite
pure state entanglement of three qubits is either equivalent to the
maximally entangled GHZ state\cite{GHZ} or to the so-called $W$ state%
\cite{Dur1}

\begin{equation}
|W\rangle =\frac 1{\sqrt{3}}(|100\rangle +|010\rangle +|001\rangle ).
\end{equation}
For the GHZ state, if one of the three qubits is traced out, the remaining state is
unentangled, which means that this state is fragile under particle losses.
Oppositely the entanglement of $W$ state is maximally robust under disposal
of any one of the three qubits \cite{Dur1}.

A natural generalization of the $W$ state to $N$ qubits and arbitrary
phases is

\begin{eqnarray}
|W_N\rangle &=&\frac 1{\sqrt{N}}(e^{i\theta _1}|1000...0\rangle +e^{i\theta
_2}|0100...0\rangle +  \nonumber \\
&&e^{i\theta _3}|0010...0\rangle +\cdot \cdot \cdot +e^{i\theta
_N}|0000...1\rangle ).  \label{eq:wn}
\end{eqnarray}
For the above state $|W_N\rangle$, the concurrences \cite{Dur1,Wootters1} between any two
qubits are all equal to $2/N$ and do not depend on the phases $\theta_i (i=1,2...N)$. This shows
that any two qubits in the $W$ state are equally entangled. Recently Koashi {\it et al.}\cite{Koashi}
shows that the maximum degree of entanglement (measured in the concurrence) between any pair
of qubits of a $N$-qubit symmetric state is $2/N$. This tight bound is achieved when the qubits
are prepared in the state $|W_N\rangle$.

The Heisenberg interaction has been used to implement quantum computer\cite{Loss}. 
It can be realized in quantum dots\cite{Loss}, nuclear
spins\cite{Kane}, electronic spins\cite{Vrijen} and optical lattices\cite
{Moelmer}. By suitable coding, the
Heisenberg interaction alone can be used for quantum computation\cite{Loss2}. 

Here we consider the quantum Heisenberg $XY$ model, which was intensively
investigated in 1960 by Lieb, Schultz, and Mattis\cite{Lieb}. Recently  Imamo\={g}lu {\it et al} 
have studied the quantum information processing using quantum dot spins and cativity QED\cite{I} and
obtained an effective interaction Hamiltonian between  two quantum dots, which is just the $XY$ Hamiltonian.
The effective Hamiltonian can be used to construct the controlled-NOT gate\cite{I}. 
The $XY$ model is also realized 
in the quantum-Hall system\cite{Hall} and in cavity QED system\cite{Zheng} 
for a quantum computer.

The $XY$ Hamiltonian is given by\cite{Lieb}
\begin{equation}
H=J\sum_{n=1}^N\left( S_n^xS_{n+1}^x+S_n^yS_{n+1}^y\right) ,  \label{eq:xy1}
\end{equation}
where $S^\alpha =\sigma ^\alpha /2$ $(\alpha =x,y,z)$ are spin 1/2
operators, $\sigma ^\alpha $ are Pauli operators, and $J>0$ is the
antiferromagnetic exchange interaction between spins. We adopt the periodic
boundary condition, i.e., $S_{N+1}^x=S_1^x,$ $S_{N+1}^y=S_1^y.$ 

One role of the $XY$ model in quantum computation is that it can be used to
construct the swap gate. The evolution operator of the corresponding
two-qubit $XY$ model is given by
\begin{equation}
U(t)=\exp \left[ -iJt(\sigma _1^x\sigma _2^x+\sigma _1^y\sigma _2^y)/2\right].
\end{equation}
Choosing $Jt=\pi /2,$ we have

\[
U\left( \frac \pi {2J}\right) |00\rangle =|00\rangle ,\text{ }U\left( \frac %
\pi {2J}\right) |11\rangle =|11\rangle, 
\]

\begin{equation}
U\left( \frac \pi {2J}\right) |01\rangle =-i|10\rangle ,\text{ }U\left( 
\frac \pi {2J}\right) |10\rangle =-i|01\rangle. 
\end{equation}
The above equation shows that the operator $U\left( \frac {\pi} {2J}\right) $ acts as
a swap gate up to a phase. Another gate $\sqrt{\text{swap}}$ which is
universal can also be constructed simply as $U\left( \frac \pi {4J}\right) .$
A swap gate can be realized by successive three C-NOT gates\cite{Youli}, while
here we only need one-time evolution of the $XY$ model. This shows that the $%
XY$ model has some potential applications in quantum computation. 

The entanglement in the ground state of the Heisenberg model has been discussed
by O'Connor and Wootters\cite{Oconnor}. Here we study the entanglement in
the $XY$ model. We first consider the generation of $W$ states in the $XY$
model. It is found that for 3 and 4 qubits, the $W$ states can be
generated at certain times. By the concept of concurrence, we study the
entanglement properties in the time evolution of the 
$XY$ model. Finally we discuss the thermal entanglement in the two-qubit $XY$
model with a magnetic field and in the anisotropic $XY$ model.

\section{Solution of the $XY$ model}

With the help of raising and lowering operators $\sigma _n^{\pm }=S_n^x\pm
iS_n^y,$ the Hamiltonian $H$ is rewritten as ($J=1)$

\begin{equation}
H={1\over 2}\sum_{n=1}^N\left( \sigma _n^{+}\sigma _{n+1}^{-}+\sigma _{n+1}^{+}\sigma
_n^{-}\right) .  \label{eq:xy2}
\end{equation}
Obviously the states with all spins down $|0\rangle ^{\otimes N}$ or all
spins up $|1\rangle ^{\otimes N}$ are eigenstates with zero eigenvalues.

The eigenvalue problem of the $XY$ model can be exactly solved by the
Jordan-Wigner transformation\cite{JW}. Here we are only interested in the time
evolution problem and in the `one particle' states ($N-1$ spins down, one
spin up), 
\begin{equation}
|k\rangle =\sum_{n=1}^Na_{k,n}\sigma _n^{+}|0\rangle ^{\otimes N}.
\end{equation}
The eigenequation is given by 
\begin{eqnarray}
H|\Psi \rangle  &=&\frac 12\sum_{n=1}^N(a_{k,n+1}+a_{k,n-1})\sigma
_n^{+}|0\rangle ^{\otimes N} \\
&=&E_k\sum_{n=1}^Na_{k,n}\sigma _n^{+}|0\rangle ^{\otimes N}.  \nonumber
\end{eqnarray}
Then the coefficients $a_{k,n}$ satisfy

\begin{equation}
\frac 12(a_{k,n+1}+a_{k,n-1})=E_ka_{k,n}.
\end{equation}
The solution of the above equation is

\begin{eqnarray}
a_{k,n} &=&\exp \left( \frac{i2\pi nk}N\right) (k=1...N), \\
E_k &=&\cos \left( \frac{2\pi k}N\right) ,\label{eq:eee}
\end{eqnarray}
where we have used the periodic boundary condition.

So the eigenvectors are given by

\begin{equation}
|k\rangle =\frac 1{\sqrt{N}}\sum_{n=1}^N\exp \left( \frac{i2\pi nk}N\right)
\sigma _n^{+}|0\rangle ^{\otimes N}
\end{equation}
which satisfy $\langle k|k^{\prime }\rangle =\delta _{kk^{\prime }}.\,$It is
interesting to see that all the eigenstates are generalized $W$ states (Eq.(\ref{eq:wn})).

Note that the $XY$ Hamiltonian $H$ commutes with the operator

\begin{equation}
Q=\sigma _x^{\otimes N}=\sigma _x^{}\otimes \sigma _x^{}\otimes ...\otimes
\sigma _x^{},
\end{equation}
then the state

\begin{equation}
|k\rangle ^{\prime }=\frac 1{\sqrt{N}}\sum_{n=1}^N\exp \left( \frac{i2\pi nk}%
N\right) \sigma _n^{-}|1\rangle ^{\otimes N}
\end{equation}
are also the eigenstates of $H$ with eigenvalues $\cos \left( 2\pi
k/N\right) .$

Now we choose the initial state of the system as $\sigma _1^{+}|0\rangle
^{\otimes N},\,$and in terms of the eigenstates $|k\rangle $, it can be expressed as

\begin{equation}
|\Psi (0)\rangle =\frac 1{\sqrt{N}}\sum_{k=1}^N\exp \left( \frac{-i2\pi k}N%
\right) |k\rangle.
\end{equation}
The state vector at time $t$ is easily obtained as

\begin{equation}
|\Psi (t)\rangle =\sum_{n=1}^Nb_n(t)\sigma _n^{+}|0\rangle ^{\otimes N},
\label{eq:psit}
\end{equation}
where 
\begin{equation}
b_n(t)=\frac 1N\sum_{k=1}^Ne^{i2\pi (n-1)k/N-it\cos \left( 2\pi k/N\right). }
\end{equation}

If we choose the initial state as $\sigma _1^{-}|1\rangle ^{\otimes N},$
then the wave vector at time $t$ will be $\sum_{n=1}^Nb_n(t)\sigma
_n^{-}|1\rangle ^{\otimes N}.$

\section{Generation of $W$ states}
From Eq.(\ref{eq:psit}), the probabilities at time $t$ for state $\sigma
_n^{+}|0\rangle ^{\otimes N}$ is obtained as

\begin{equation}
P(n,N,t)=|b_n(t)|^2.
\end{equation}

For $N=2$, it is easy to see that the probability $P(1,2,t)=\cos ^2t$ , $%
P(2,2,t)=\sin ^2t.\,$The state vector at time $t$ is

\begin{equation}
|\Psi (t)\rangle =\cos t|10\rangle -i\sin t|01\rangle.
\end{equation}
When $t=\pi /4,$ the above state is the maximally entangled state.

Now we consider the case $N=3.$ The probabilities are analytically obtained
as

\begin{eqnarray}
P(1,3,t) &=&\frac 19\left[ 5+4\cos \left( \frac 32t\right) \right] , 
\nonumber \\
P(2,3,t) &=&P(3,3,t)=\frac 19\left[ 2-2\cos \left( \frac 32t\right) \right] .
\label{eq:p1}
\end{eqnarray}
Fig.1(a) gives a plot of the probabilities versus time. It is clear that
there exist some cross points of the probabilities. At these special times
the probabilities $P(n,3,t)$ are all equal to 1/3, which indicates the $W$
states are generated. From Eq.(\ref{eq:p1}), we see that if the time $t$
satisfies the equation

\begin{equation}
\cos \left( \frac 32t\right) =-\frac 12,  \label{eq:t}
\end{equation}
the probabilities are same. The solution of Eq.(\ref{eq:t}) is

\begin{eqnarray}
t_n &=&\frac{4\pi }9+\frac{4n\pi }3,  \nonumber \\
t_n^{\prime } &=&\frac{8\pi }9+\frac{4n\pi }3(n=0,1,2,...).
\end{eqnarray}
Explicitly at these time points, the corresponding state vectors are

\begin{eqnarray}
|\Psi (t_n)\rangle &=&\frac 1{\sqrt{3}}\left( |100\rangle +e^{\frac{-i2\pi }3%
}|010\rangle +e^{\frac{-i2\pi }3}|001\rangle \right) ,  \nonumber \\
|\Psi (t_n^{\prime })\rangle &=&\frac 1{\sqrt{3}}\left( |100\rangle +e^{%
\frac{i2\pi }3}|010\rangle +e^{\frac{i2\pi }3}|001\rangle \right)
\end{eqnarray}
which are the generalized $W$ state for $N=3.$

\begin{figure}[tbh]
\begin{center}
\epsfxsize=8cm
\epsffile{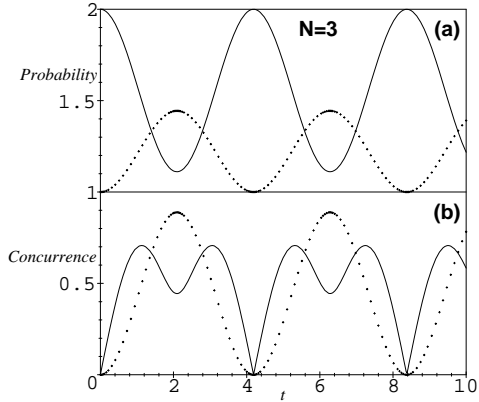}
\end{center}
\caption{{{\protect\small Time evolution of the probabilities and
concurrences for $N=3$. (a)The probablity plus 1 for $n=1$ (solid line) and $%
n=2, n=3$ (dotted line); (b)The concurrences $C_{12}(t), C_{13}(t)$(solid
line) and $C_{23}(t)$}} (dotted line).}
\label{fig1}
\end{figure}

For the case $N=4,$ the probabilities are given by

\begin{eqnarray}
P(1,4,t) &=&\cos ^4\left( \frac t2\right) ,\text{ }P(3,4,t)=\sin ^4\left( 
\frac t2\right) ,  \nonumber \\
P(2,4,t) &=&P(4,4,t)=\frac 14\sin ^2t,
\end{eqnarray}

As seen from Fig.2(a), there also exists some cross points, which indicates
the 4-qubit $W$ states are generated. The probabilities are same when 
\begin{eqnarray}
t_n &=&\frac \pi 2+2n\pi , \\
t_n^{\prime } &=&\frac{3\pi }2+2n\pi (n=0,1,2,...)  \nonumber
\end{eqnarray}
Explicitly the 4-qubit $W$ states are\newline
\begin{eqnarray}
|\Psi (t_n)\rangle  &=&\frac 12\left( |1000\rangle -i|0100\rangle
-|0010\rangle -i|0001\rangle \right) ,  \nonumber \\
|\Psi (t_n^{\prime })\rangle  &=&\frac 12\left( |1000\rangle +i|0100\rangle
-|0010\rangle +i|0001\rangle \right) .  \nonumber \\
&&
\end{eqnarray}

\begin{figure}[tbh]
\begin{center}
\epsfxsize=8cm
\epsffile{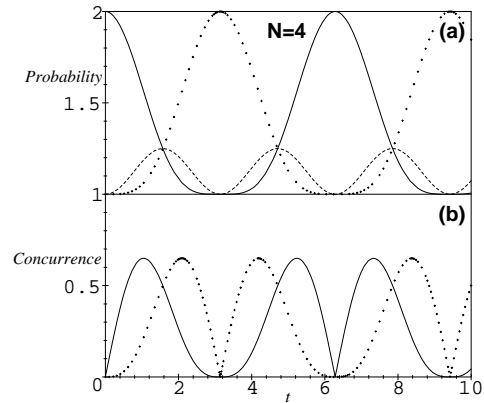}
\end{center}

\caption{{{\protect\small Time evolution of the probabilities and
concurrences for $N=4$. (a)The probablity plus 1 for $n=1$ (solid line), $n=3
$ (dotted line) and $n=2, n=4$ (dashed line); (b)The concurrences $C_{12}(t)$
(solid line) and $C_{23}(t)$(dotted line).}}}
\label{fig2}
\end{figure}

Can we generate $W$ states for more than 4 qubits in the $XY$ model?
Fig.3(a) shows that there is no cross points for $N=5$ . Further numerical
calculations for long time and large $N$ show no evidence that there exist
some times at which the $W$ states can be generated.

\begin{figure}[tbh]
\begin{center}
\epsfxsize=8cm
\epsffile{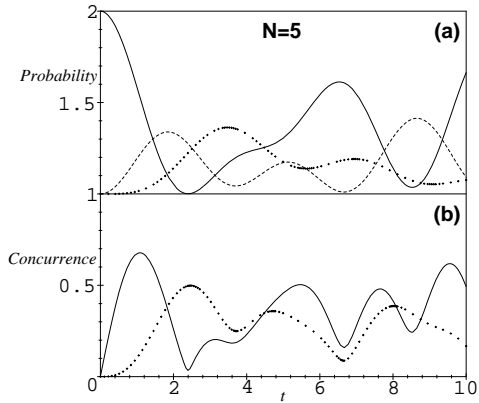}
\end{center}

\caption{{{\protect\small Time evolution of the probabilities and
concurrences for $N=5$. (a)The probablity plus 1 for $n=1$ (solid line), $%
n=3, n=4$ (dotted line) and $n=2, n=5$ (dashed line);(b)The concurrences $%
C_{12}(t)$ (solid line) and $C_{23}(t)$(dotted line). }}}
\label{fig3}
\end{figure}

We see that the $W$ states appear periodically for 3 and 4 qubits. In order
that a certain state occurs periodically in a system, a necessary condition is that
the ratio of any two frequencies available in the system is a rational number.
From Eq.(\ref{eq:eee}) it is easy to check that the necessary condition is satisfied 
for 2, 3, 4, and 6 qubits. For 6 qubits the corresponding probabilities evolve periodically
with time. The numerical calculations show that there exists no cross points, i.e., 
we can not creat 6-qubit $W$ state. However some states close to the $W$ state will occur 
repeatedly and these states may be used for quantum computation.

The 3-qubit and 4-qubit $W$ states are readily generated by only
one-time evolution of the $XY$ system. This idea is similar to the
concurrent quantum computation\cite{cqc} in which some functions of
computation are realized by only one-time evolution of  multi-qubit
interaction systems.

The $W$ entangled states can be generated by other methods, such as coupling $N$
spins with a quantized electromagnetic field. However here we only use the
interaction of $N$ spins themselves and do not need to introduce additional
degree of freedoms.

\section{Time evolution of entanglement}

We first briefly review the definition of concurrence\cite{Wootters1}. Let $\rho
_{12}$ be the density matrix of a pair of qubits $1$ and $2.$ The density
matrix can be either pure or mixed. The concurrence corresponding to the
density matrix is defined as

\begin{equation}
C_{12}=\max \left\{ \lambda _1-\lambda _2-\lambda _3-\lambda _4,0\right\} 
\label{eq:c1}
\end{equation}
where the quantities $\lambda _1\ge \lambda _2\ge \lambda _3\ge $ $\lambda _4
$ are the square roots of the eigenvalues of the operator 
\begin{equation}
\varrho_{12}=\rho _{12}(\sigma _y\otimes \sigma _y)\rho _{12}^{*}(\sigma _y\otimes \sigma
_y).  \label{eq:c2}
\end{equation}
The nonzero concurrence implies that the qubits 1 and 2 are entangled.
The concurrence $C_{12}=0$ corresponds to an unentangled state and $C_{12}=1$
corresponds to a maximally entangled state.

We consider the entanglement in the state $|\Psi (t)\rangle $(\ref{eq:psit}%
). By direct calculations, the concurrence between any two qubits $i$ and 
$j$ are simply obtained as
\begin{equation}
C_{ij}(t)=2|b_i(t)b_j(t)|.  \label{eq:c}
\end{equation}
The numerical results for the concurrence are shown in Fig.1(b), Fig.2(b)
and Fig.3(b).

For $N=3,$ Fig.1(b) shows that the entanglement is periodic with period $%
4\pi /3.$ At times $4n\pi /3(n=1,2,3,...)$, the state vectors are
disentangled$\,$and become the state $|100\rangle $ up to a phase. The
concurrences of $C_{12}(t)$ and $C_{13}(t)$ are same, and have two maximum
points in one period, while the concurrence $C_{13}(t)$ has only one maximum
point. Fig.2(b) shows the concurrences for $N=4.$ They are periodic with
period $2\pi .$ In one period there are two unentanglement points, $t=\pi
,2\pi .$ For both concurrences $C_{12}(t)$ and $C_{23}(t)$, there are two
maximum points in one period. If we choose large $N$ (see Fig.3(b) for $N=5$%
), there exists no exact periodicity for the entanglements of two qubits. From
the time evolution of the concurrences we can see clearly when the
system becomes disentangled and when the system maximally entangled.

\section{Thermal entanglement}

Recently the concept of thermal entanglement was introduced and studied
within one-dimensional isotropic Heisenberg model\cite{Arnesen}. Here we study this
kind of entanglement within both the isotropic $XY$ model with a magnetic
field and the anisotropic $XY$ model.

\subsection{Isotropic $XY$ model with a magnetic field}

We consider the two-qubit isotropic antiferromagnetic $XY$ model in a constant external
magnetic field $B,$

\begin{equation}
H=\frac B2(\sigma _1^z+\sigma _2^z)+J\left( \sigma _1^{+}\sigma
_2^{-}+\sigma _2^{+}\sigma _1^{-}\right) .
\end{equation}
The eigenvalues and eigenvectors of $H$ are easily obtained as

\begin{eqnarray}
H|00\rangle  &=&-B|00\rangle ,H|11\rangle =B|11\rangle ,  \nonumber \\
H|\Psi ^{\pm }\rangle  &=&\pm J|\Psi ^{\pm }\rangle ,  \label{eq:evals}
\end{eqnarray}
where $|\Psi ^{\pm }\rangle =\frac 1{\sqrt{2}}(|01\rangle \pm |10\rangle )$
are maximally entangled states.

The state of the system at thermal equilibrium is $\rho (T)=\exp \left( -%
\frac H{kT}\right) /Z,$ where $Z=$Tr$\left[ \exp \left( -\frac H{kT}\right)
\right] $ is the partition function and $k$ is the Boltzmann's constant.
As $\rho (T)$ represents a thermal state, the entanglement in the state is
called thermal entanglement\cite{Arnesen}.

In the standard basis, $\left\{ |00\rangle ,|01\rangle ,|10\rangle
,|11\rangle \right\} ,$ the density matrix $\rho (T)$ is written as ($k=1$)

\begin{eqnarray}
\rho (T) &=&\frac 1{2(\cosh \frac JT+\cosh \frac BT)}  \label{eq:rho} \\
&&\times \left( 
\begin{array}{llll}
e^{-\frac BT} & 0 & 0 & 0 \\ 
0 & \cosh \frac JT & -\sinh \frac JT & 0 \\ 
0 & -\sinh \frac JT & \cosh \frac JT & 0 \\ 
0 & 0 & 0 & e^{\frac BT}
\end{array}
\right)  \nonumber
\end{eqnarray}
From Eqs.(\ref{eq:c1}),(\ref{eq:c2}) and (\ref{eq:rho}), the concurrence
is given by

\begin{equation}
C=\max \left( \frac{\sinh \frac JT-1}{\cosh \frac JT+\cosh \frac BT}%
,0\right) .  \label{eq:cc}
\end{equation}

Then we know $C=0$ if $\sinh \frac JT\leq 1,\,$i.e., there is a critical
temperature

\begin{equation}
T_c=\frac J{\arcsin \text{h}(1)}\approx 1.1346J,
\end{equation}
the entanglement vanishes for $T\geq T_c.$ It is interesting to see that the
critical temperature is independent on the magnetic field $B.$

For $B=0,$ the maximally entangled state $|\Psi ^{-}\rangle $ is the ground
state with eigenvalue $-J.$ Then the maximum entanglement is at $T=0,$ i.e., 
$C=1.\,$As $T$ increases, the concurrence decreases as seen from Fig.4 due
to the mixing of other states with the maximally entangled state. For a high
value of $B$ (say $B=1.2$), the state $|00\rangle \,$becomes the ground
state, which means there is no entanglement at $T=0.$ However by increasing $%
T,$ the maximally entangled states $|\Psi ^{\pm }\rangle $ will mix with the state $%
|00\rangle ,$ which makes the entanglement increase (see Fig.4).

\begin{figure}[tbh]
\begin{center}
\epsfxsize=8cm
\epsffile{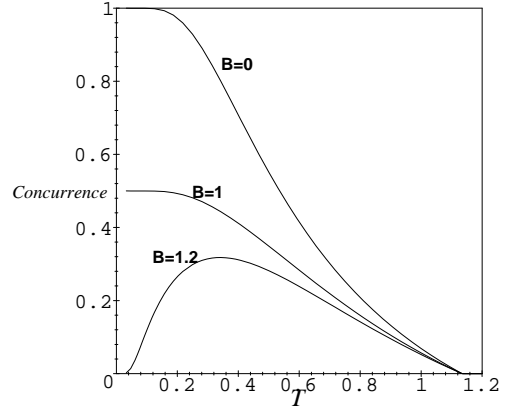}
\end{center}
\caption{{\protect\small The concurrences versus temperature for different
magnetic fields. The parameter $J$ is set to one.}}
\label{fig4}
\end{figure}

From Fig.5 we see that there is a evidence of phase transition for small
temperature by increasing magnetic field. Now we do the limit $T\rightarrow
0\,$on the concurrence (\ref{eq:cc}), we obtain 
\begin{eqnarray}
\lim_{T\rightarrow 0}C &=&1\text{ for }B<J,  \nonumber \\
\lim_{T\rightarrow 0}C &=&\frac 12\text{ for }B=J,  \nonumber \\
\lim_{T\rightarrow 0}C &=&0\text{ for }B>J.
\end{eqnarray}
So we can see that at $T=0,$ the entanglement vanishes as $B$ crosses the
critical value $J.$ This is easily understand since we see that if $B>J,$
the ground state will be the unentangled state $|00\rangle .$ This special
point $T=0,B=J,$ at which entanglement becomes a nonanalytic function of $B,$
is the point of quantum phase transition\cite{QPT}.

\begin{figure}[tbh]
\begin{center}
\epsfxsize=8cm
\epsffile{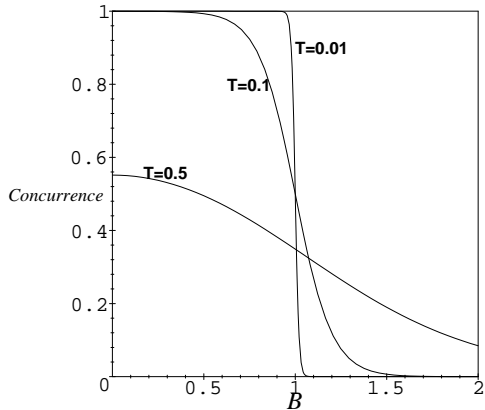}
\end{center}

\caption{{\protect\small The concurrences versus magnetic field B for
different temperatures. The parameter $J$ is set to one.}}
\label{fig5}
\end{figure}

It should be pointed out that the results of thermal entanglement in the
present isotropic $XY$ model is qualitatively the same as but quantitatively
different from that in the isotropic Heisenberg model\cite{Arnesen}. An important conclusion
is that the
concurrences are the same for both positive $J$ and negative $J$ in the $XY$
model. That is to say, the entanglement exists for both antiferromagnetic
and ferromagnetic cases. In contrary to this, for the case of two-qubit
Heisenberg model, no thermal entanglement exists for the ferromagnetic case.

\subsection{Anisotropic $XY$ model}

Now we consider the two-qubit anisotropic antiferromagnetic $XY$ model which is described by
the Hamiltonian\cite{Lieb}

\begin{eqnarray}
H_a &=&\frac J2\left[ (1+\gamma )\sigma _1^x\sigma _2^x+(1-\gamma )\sigma
_1^y\sigma _2^y\right] , \\
&=&J\left( \sigma _1^{+}\sigma _2^{-}+\sigma _2^{+}\sigma _1^{-}\right)
+J\gamma \left( \sigma _1^{+}\sigma _2^{+}+\sigma _2^{-}\sigma _1^{-}\right)
.  \nonumber
\end{eqnarray}
where $\gamma $ is the anisotropic parameter. Obviously the eigenvalues and
eigenvectors of the Hamiltonian $H_a$ is given by $H_a|\Psi ^{\pm }\rangle
=\pm J|\Psi ^{\pm }\rangle $ and $H_a|\Phi ^{\pm }\rangle =\pm J\gamma |\Phi
^{\pm }\rangle ,$ where $|\Phi ^{\pm }\rangle =\frac 1{\sqrt{2}}(|00\rangle
\pm |11\rangle ).$ Then the four maximally entangled Bell states are the
eigenstates of the Hamiltonian $H_a.$ Although the anisotropic
parameter can be arbitrary, we restrict ourselves on $0\leq \gamma \leq 1.$
The parameter $\gamma =0$ and 1 correspond to the isotropic $XY$ model and
Ising model respectively. Thus the anisotropic $XY$ model can be considered
as a interpolating Hamiltonian between the isotropic $XY$ model and the Ising model.
The anisotropic parameter $\gamma$ controls the interpolation.

The density matrix $\rho (T)$ in the standard basis is given by

\begin{center}
$\rho (T)=\frac 1{2(\cosh \frac JT+\cosh \frac{J\gamma }T)}$

$\times \left( 
\begin{array}{llll}
\cosh \frac{J\gamma }T & 0 & 0 & -\sinh \frac{J\gamma }T \\ 
0 & \cosh \frac JT & -\sinh \frac JT & 0 \\ 
0 & -\sinh \frac JT & \cosh \frac JT & 0 \\ 
-\sinh \frac{J\gamma }T & 0 & 0 & \cosh \frac{J\gamma }T
\end{array}
\right) $
\end{center}

\begin{equation}
\end{equation}
The square root of the eigenvalues of the operator $\varrho_{12}$ are $\frac{e^{\pm
J/T}}{2(\cosh \frac JT+\cosh \frac{J\gamma }T)}$ and $\frac{e^{\pm J\gamma
/T}}{2(\cosh \frac JT+\cosh \frac{J\gamma }T)}.\,$ Then from Eq.(\ref{eq:c1}), the 
concurrence is given by

\begin{equation}
C=\max \left( \frac{\sinh \frac JT-\cosh \frac{J\gamma }T}{\cosh \frac JT%
+\cosh \frac{J\gamma }T},0\right)   \label{eq:caxy}
\end{equation}
As we expected Eq. (\ref{eq:caxy}) reduces to Eq. (\ref{eq:cc}) with $B=0$
when $\gamma =0.$ When $\gamma =1,$ the concurrence $C=0,$ which indicates
that no thermal entanglement appears in the two-qubit Ising model. In this
anisotropic model, the concurrences are the same for both positive $J$ and
negative $J$ , i.e, the thermal entanglement is the same for the
antiferromagnetic and ferromagnetic cases. The critical temperature $T_{c%
\text{ }}$is determined by the nonlinear equation 
\[
\sinh \frac JT=\cosh \frac{J\gamma }T,
\]
which can be solved numerically.

In Fig.6 we give a plot of the concurrence as a function of temperature $T$
for different anisotropic parameters. At zero temperature the concurrence
is 1 since no matter what the sign of $J$ is and what the values of $\gamma $
are, the ground state is one of the Bell states, the maximally entangled state.
The concurrence monotonically decreases with the increase of temperature
until it reaches the critical value of $T$ and becomes zero. The numerical
calculations also show that the critical temperature decreases as the
anisotropic parameter increases from 0 to 1.

\begin{figure}[tbh]
\begin{center}
\epsfxsize=8cm
\epsffile{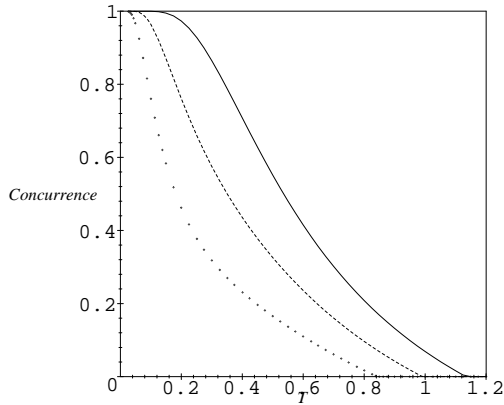}
\end{center}
\caption{{\protect\small The concurrences versus temperature for different
aniosotropic parameters: $\gamma=0$ (solid line), $\gamma=0.6$ (dashed line)
and $\gamma=0.8$ (dotted line). The parameter $J$ is set to one.}}
\label{fig6}
\end{figure}

\section{Conclusions}

In conclusion, we have presented some interesting results in the simple $XY$
model. First, we can use $XY$ interaction to generate the 3-qubit and 4-qubit $W$
entangled states. Second, we see that the time evolution of entanglement are
periodic for 2, 3, 4 and 6 qubits, and there is no exact periodicity
for large $N$. At some special points the states becomes
disentangled. Finally we study the thermal entanglement within a two-qubit isotropic
$XY$ model with a magnetic field and an anisotropic $XY$ model, and
find that the thermal entanglement exists for both ferromagnetic and
antiferromagnetic cases. Even in the simple model we see some evidence of
the quantum phase transition.

The entanglement is not completely determined by the partition function,
i.e., by the usual quantum statistical physics. It is a good challenge to
study the entanglement in multi-qubit quantum spin models.

\acknowledgments

The author thanks Klaus M\o lmer and Anders S\o rensen for many valuable
discussions and thanks for the referee's valuable comments. 
This work is supported by the Information Society 
Technologies Programme IST-1999-11053, EQUIP, action line 6-2-1.

\end{document}